# Single photon detection performance of highly disordered NbTiN thin films

Ruoyan Ma[1,2,3], Rui Shu[4], Xingyu Zhang[1,2], Aobo Yu[1,2,3], Huang Jia[1,2], You Xiao[1,2,3], Huiqin Yu[1,2], Xiaoyu Liu[1,2], Hao Li[1,2,3], Per Eklund[4], Xiaofu Zhang[1,2,3,*], Lixing You[1,2,3,*]

[1]State Key Laboratory of Functional Materials for Informatics, Shanghai Institute of Microsystem and Information Technology, Chinese Academy of Sciences (CAS), Shanghai 200050, China

[2]University of Chinese Academy of Sciences, Center of Materials Science and Optoelectronics Engineering, Beijing, China, 100049

[3]CAS Center for Excellence in Superconducting Electronics (CENSE), Shanghai 200050, China

[4]Thin Film Physics Division, Department of Physics Chemistry, and Biology (IFM), Linköping University, Linköping SE-581 83, Sweden

E-mail: zhangxf@mail.sim.ac.cn and lxyou@mial.sim.ac.cn

**Abstract**

We experimentally investigated the detection performance of highly disordered $Nb_xTi_{1-x}N$ based superconducting nanowire single photon detectors (SNSPDs). The dependence on the composition of the transition temperature $T_c$ for $Nb_xTi_{1-x}N$ films show a dome-like behavior on the Nb content, with a maximal $T_c$ at $x_{Nb}\sim0.65$, and the $Nb_{0.65}Ti_{0.35}N$ films also combine relatively large sheet resistance and intermediate residual resistivity ratio. Moreover, 60-nm-wide and 7-nm-thick $Nb_{0.65}Ti_{0.35}N$ nanowires show a switching current as high as 14.5 µA, and saturated intrinsic detection efficiency with a plateau of more than 2 µA at 2.4 K. Finally, the corresponding SNSPDs on an alternative $SiO_2/Ta_2O_5$ dielectric mirror showed a system detection efficiency of approximately 92% for 1550 nm photons, and the timing jitter is around 26 ps. Our results demonstrate that the highly disordered $Nb_xTi_{1-x}N$ films are promising for fabricating SNSPDs for near- and middle-infrared single photons with high detection efficiency and low timing jitter.

**Introduction**

Superconducting nanowire single photon detectors (SNSPDs) with high system detection efficiency, low timing jitter, low dark counts, and high counting rate have widely been applied in broad fields for both quantum and classical realms, such as quantum optics [1-3], quantum information processing [4-7]. Especially for telecom-wavelength photons at 1550 nm, the SNSPDs have shown high system detection efficiency (SDE) up to 98% for both disordered bilayer-NbN and amorphous MoSi [8,9]. In these two systems, however, the operation temperature and maximum applicable bias current are limited due to the suppressed superconductivity, which subsequently limits the timing performance of SNSPDs. To further optimize the detection performance of SNSPDs, a significant improvement in intrinsic detection performance was recently realized in the NbTiN-based SNSPDs [10-12]. By optionally doping Ti into NbN films, it was found that a SDE of ~90% at 1550 nm with good timing performance was realized in NbTiN-SNSPDs operated in Gifford–McMahon cryocoolers [11].

In NbTiN films, the additional Ti atoms are also able to largely prevent the formation of vacancy defect and thus improve the quality of crystal grains, which in turn enhances the superconducting and electrical properties of granular NbTiN films [13,14]. As a result, NbTiN thin films can be deposited on various substrates without intentionally heating the substrate [15]. However, due to the introduction of Ti, the composition and structure of ternary NbTiN compound are more complexly dependent on the deposition conditions during the sputtering process, which makes the optimization of NbTiN thin films towards SNSPD fabrication more complicated. Recently, it has been experimentally demonstrated that the intrinsic detectionefficiency (IDE) of SNSPDs may related with the sheet resistance $R_s$ and the residual resistivity ratio (*RRR*) [16,17]. The saturation plateau of SNSPDs was found to decrease with the decreasing $R_s$, namely, the disorder level of superconducting thin films. In highly disordered superconducting thin films, in which the electron-electron (e-e) interaction time $\tau_{ee}$ is much shorter than the electron-phonon interaction time $\tau_{eph}$ [18], the non-equilibrium quasiparticles relax mainly through the e-e interaction [19-21]. Subsequently, a higher portion of photon energy will be transferred into the electronic system instead of the phonon system, which drives more Cooper pairs into quasiparticles and leads to a higher intrinsic quantum efficiency [19]. Beyond the sheet resistance or disorder level of superconducting thin films, it has also been demonstrated that the detection performance of SNSPDs was also related to the *RRR* of superconducting films [17], where the saturation of *IDE* increases in films with a lower value of *RRR*. As a result of $R_s$ and *RRR* dependence of IDE, to deposit disordered NbTiN thin films with relatively large $R_s$ and low *RRR* would be highly suitable for SNSPD fabrications.

Here, we investigate the superconducting properties of highly disordered $Nb_xTi_{1-x}N$ films with different Nb content, where $x$ is determined by atomic percentage ratio between Nb and total metals. From the composition dependence on superconducting critical temperature $T_c$, $R_s$, and *RRR*, we figured out the optimal $Nb_{0.65}Ti_{0.35}N$ films for SNSPDs fabrications among our disordered $Nb_xTi_{1-x}N$ films. We then measured intrinsic detection performance of $Nb_{0.65}Ti_{0.35}N$-SNSPDs on SiO₂/Si substrate. Finally, by fabricating $Nb_{0.65}Ti_{0.35}N$-SNSPDs on an alternative SiO₂/Ta₂O₅ dielectric mirror, a SDE up to 92% and timing jitter of 26 ps are simultaneously obtained at a temperature of 2.4 K.

**Experimental details**
The $Nb_xTi_{1-x}N$ films were deposited by reactive direct-current magnetron sputtering with elemental Nb (purity: 99.99%) and Ti (purity: 99.99%) targets in a high vacuum chamber (base pressure ~$8 \times 10^{-8}$ Pa). During the deposition, a gas mixture of 10 sccm Ar and 2 sccm N₂ was simultaneously introduced to each target, corresponding to a deposition pressure around 0.25 Pa. The NbTiN films were deposited on a constant current mode. The deposition rate was determined by a thick film (>200 nm) with relatively long deposition time. The nominal thickness *d* of the resulting thin films was inferred from the predetermined deposition rate, and confirmed by both atomic force microscope and transmission electron microscopy. To obtain uniform NbTiN films for device fabrications, the deposition rate was kept relatively low, around 0.04 nm/s. To investigate the physical properties and photon detection performance of the

NbTiN films, they have been deposited on both thermally oxidized Si substrates and alternative $SiO_2/Ta_2O_5$ dielectric mirror.

The composition of the films was determined by Rutherford backscattering spectroscopy (RBS) by using 2-MeV $^4He^+$ ions beam on $Nb_xTi_{1-x}N$ films with a thickness around 150 nm. Backscattered ions were detected at a scattering angle of 170°. Channeling effects in the Si substrates and potentially the textured films were minimized by adjusting the equilibrium incidence angle to 5° with respect to the surface normal and perform multiple-small-random-angular movements within a range of 2° during data acquisition. Atomic concentrations were extracted from the spectra using the SIMNRA simulation program, which are summarized in table 1. All concentration values are presented in atomic ratio and accurate to within ±0.5 at. %. The crystal structure was evaluated by X-ray diffraction (XRD) measurements with a PANalytical X'Pert PRO diffractometer in a Bragg-Brentano geometry. Transmission electron microscopy (TEM) characterization was conducted with a FEI Tecnai G2 TF20 UT instrument with a field emission gun operated at 200 kV. The specimen for TEM examination were made by mechanical grinding, followed by Ar+-ion milling with a Gatan precision ion polishing system. Initially, the incident angle and energy of the $Ar^+$ ions were set at 5.5° and 4.0 keV, respectively. During the final step, the $Ar^+$ energy was decreased to 2.5 keV.

Table 1 Composition and Physical properties of $Nb_xTi_{1-x}N$ films

| $x_{Nb}$ | Nb (at. %) | Ti (at. %) | N (at. %) | $T_c$ (K) | $\rho_n$ (μΩ·cm) | $R_s$ (Ω) | RRR |
|---|---|---|---|---|---|---|---|
| 0.44 | 18.0 | 23.0 | 59.0 | 11.11 | 823 | 61.4 | 0.89 |
| 0.54 | 22.0 | 19.0 | 59.0 | 11.71 | 743 | 53.1 | 0.87 |
| 0.59 | 24.0 | 17.0 | 59.0 | 11.86 | 704 | 47.9 | 0.84 |
| 0.65 | 27.5 | 14.5 | 58.0 | 12 | 664 | 46.1 | 0.79 |
| 0.72 | 31.0 | 12.0 | 57.0 | 11.75 | 618 | 42.9 | 0.75 |
| 0.84 | 36.0 | 7.0 | 57.0 | 11.35 | 569 | 38.2 | 0.7 |

The as-grown NbTiN films were then fabricated into microbridges by photo-lithography and reactive ion etching in the $CF_4$ plasma. The transport measurements were subsequently carried out in a physical property measurement system (PPMS) from *Quantum Design* under perpendicular magnetic fields up to 9 T. The microbridges were wire-connected in a four-point probing configuration. The resistivity $\rho$ or sheet resistance $R_s$ was calculated from the measured total resistance and the bridge geometry. Finally, the deposited NbTiN films were patterned into superconducting nanowires via standard electron beam lithography.

**Results and discussions**
Figure 1(a) shows X-ray diffractograms of the $Nb_xTi_{1-x}N$ films as function of $x$ for films with thickness around 150 nm. The diffractogram of all the films shows a dominant peak located near 35.5°, which could be attributed to 111 reflections of a NaCl-type *fcc* structure. When the alloying Nb content $x$ increases from 0.44 to 0.84, the 111 reflections are shifted to lower 2θ angles (larger d spacings). In figure 1b, the lattice constant $a$ was calculated based on 111 reflections. $a$ increases as a function of $x$ and obeys the trend estimated by Vegard's law. Moreover, plan-view transmission electron microscopy (TEM) was performed (Figure 1c), showing a typical faceted column tops of the early-transition-metal nitrides (such as TiN) [22]. The selected-area electron diffraction (SAED) pattern (Figure 1d) shows a diffraction pattern of cubic *fcc* structure, which is consistent with the XRD pattern in Figure 1a. From the nitrogen composition and the crystal structure, the $Nb_xTi_{1-x}N$ films in this study present a similar $\delta$ phase as compared with NbN [23-25].

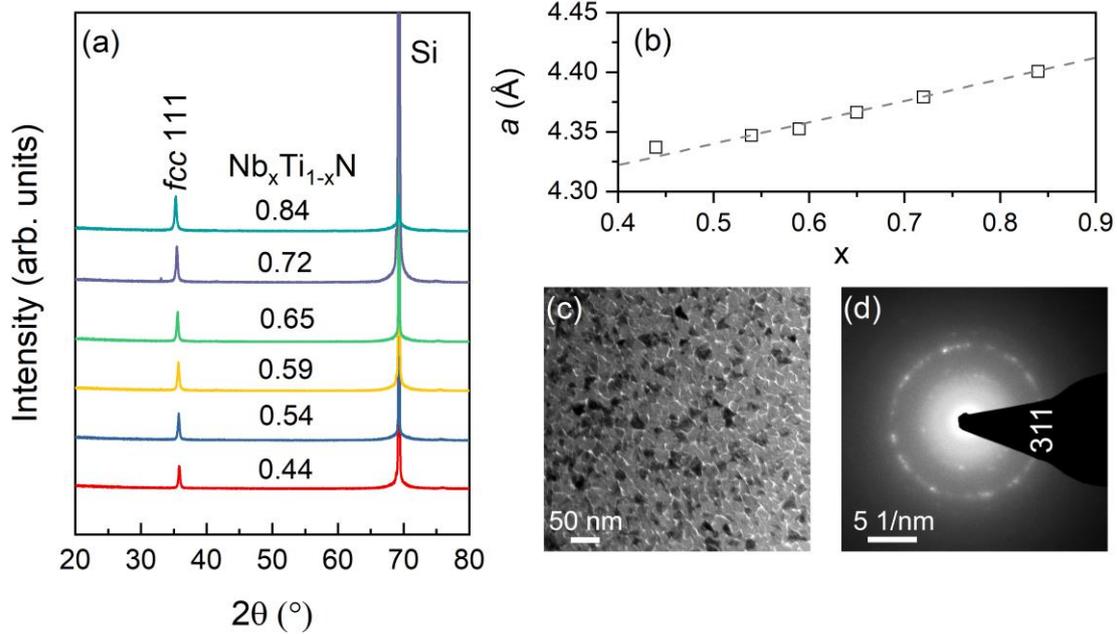

Figure 1. (a) X-ray diffractograms of the $Nb_xTi_{1-x}N$ films with x varied from 0.44 to 0.84. (b) Lattice constant a, calculated based on 111 reflections as a function of x, showing a consistent with the values estimated by Vegard's law. (c) Plan-view TEM for the $Nb_{0.59}Ti_{0.41}N$ film. (d) SAED pattern from region c, revealing a polycrystalline cubic structure.

To reveal the composition dependence of superconducting properties of $Nb_xTi_xN$ films, we here firstly investigated the normal to superconducting transitions for both bulk and thin films. In figure 2(a), we first measured the resistivity as a function of temperature for all films, in which all NbTiN films show a weak insulating behavior above 20 K in the normal state. To compare the superconducting properties, the temperature dependence of resistivity is therefore normalized to the maximal resistivity at $T \sim 20$ K. The normal state resistivity (at 20 K) monotonically increases with the decreasing Nb content, ranging from 813 $\mu\Omega \cdot cm$ for $x = 0.85$ to 924 $\mu\Omega \cdot cm$ for $x = 0.44$. Basically, the resistivity of these NbTiN films are much higher than those from previous publications [10,13,14,26,27], representing a more disordered NbTiN thin film system. The ratio $\rho_{300}/\rho_{20}$ (RRR), on the contrary, increase from 0.7 for $x = 0.85$ to 0.89 for $x = 0.44$. By fitting the normal to superconducting transition with the three-dimensional (3D) fluctuating mechanism [28-30], the superconducting critical temperature $T_c$ is obtained and summarized in figure. 2(b) and table 1. Different from the previous publications, in which the transition temperature decreases monotonically with the reducing Nb content [10], our highly disordered thick $Nb_xTi_xN$ films show a dome-like $T_c$ dependence on $x_{Nb}$. From the composition dependence of $R_s$ and RRR summarized in table 1, we conclude that thin

$Nb_{0.65}Ti_{0.35}N$ films would be highly suitable for SNSPD fabrications, which also holds the highest $T_c$ of 12 K among all investigated films.

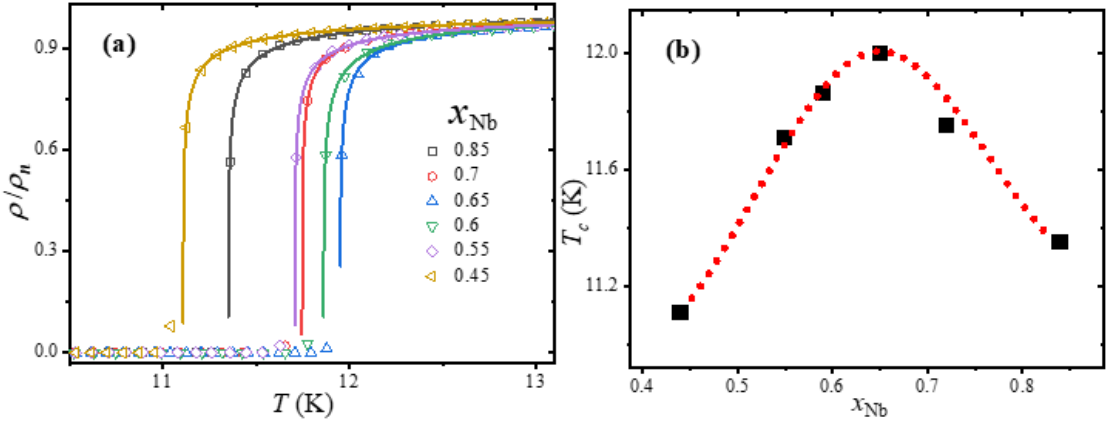

Figure 2. (a) Normalized resistivity as a function of temperature for bulk $Nb_xTi_{1-x}N$ films near the transition region. The solid lines are fittings to the 3D superconducting fluctuating mechanism. (b) The $T_c(x_{Nb})$ dependence for thick films. The red dotted line is a guide to the eye.

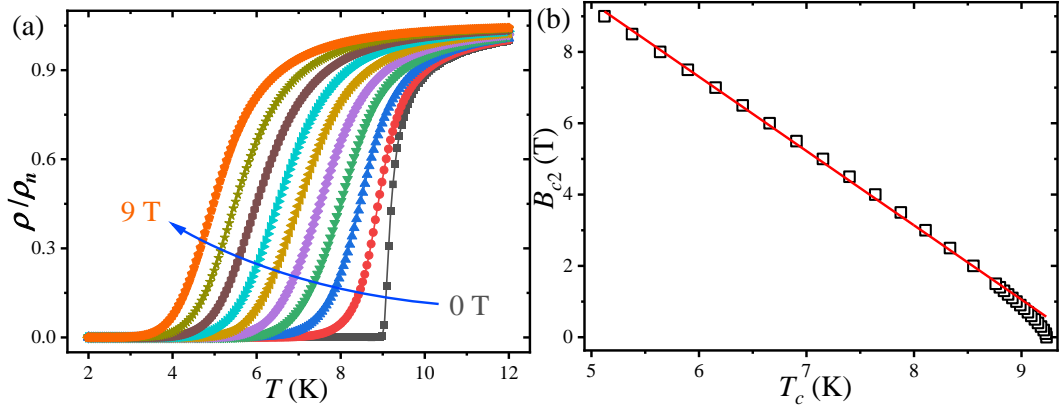

Figure 3. (a) Magnetic field-dependent normal to superconducting transitions from 0 to 9 T for a 7-nm-thick $Nb_{0.65}Ti_{0.35}N$ microbridge. The bridge width and the length are 6 μm and 100 μm, respectively. (b) Magnetic field dependence of superconducting critical temperature.

To guarantee good photon absorption efficiency and excellent intrinsic detection performance of superconducting nanowires, we here mainly focus on 7-nm-thick films. To investigate the superconducting properties of 7-nm-thick $Nb_{0.65}Ti_{0.35}N$ films, we firstly deposited films on $SiO_2$/Si substrate, and then patterned them into microbridges. Figure 3(a) shows the normal to superconducting transitions in magnetic fields from 0 to 9 T. With the increasing fields, the transition regions broaden slightly. By defining $T_c$ at temperatures where resistivity drops to half of the normal state resistivity, the resulting $T_c$ gradually decreases from 9.23 K to 5.12 K, and the field dependence of $T_c$ is plotted in figure 3(b). By linearly fitting the $T_c(B)$ dependence, a zero-temperature upper critical field of $B_{c2}(0) = 13.7$ T can be obtained [31,32]. From Ginzburg–Landau (GL) theory, the zero-temperature GL coherence length can be estimated by $\xi_{GL}(0) = [\Phi_0/2\pi B_{c2}(0)]^{0.5} = 4.9$ nm [30]. Moreover, the diffusion coefficient of the normal state electrons can then be estimated by $D_e = -1.097(dB_{c2}/dT)^{-1} = 0.52$ cm$^2$/s [33].

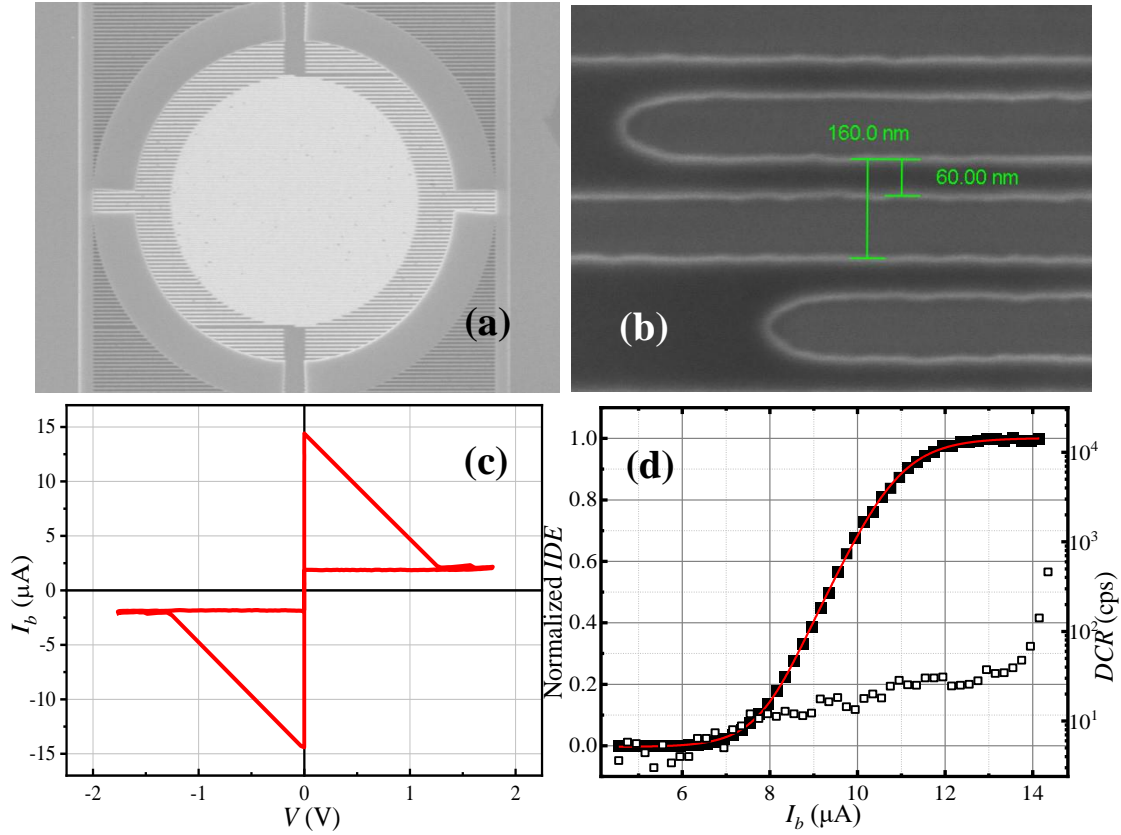

Figure 4. (a) SEM image of the fabricated NbTiN-SNSPD on SiO₂/Si substrate. The sensitive area is 18 μm in diameter. (b) SEM image of the fabricated nanowires and the rounded corners connecting meander nanowires. (c) the current voltage characteristics of the fabricated NbTiN-SNSPD. (d) Bias current dependences of IDE and DCR. The red solid line is a sigmoid curve fitting to IDE.

To determine the intrinsic detection performance of 7-nm-thick $Nb_{0.65}Ti_{0.35}N$ films, we first fabricated SNSPDs on SiO₂/Si substrates, and systematically investigated their detection performance. Figure 4(a) depicts an overall scanning electron microscope (SEM) image of the resulting SNSPDs, which covers a sensitive area of 18 μm. In figure 4(b), we present an SEM image of the corner of the devices. The nanowire width and pitch are 60 and 160 nm, respectively, leading to a filling factor of 37.5%. The 180-degree turnaround is intentionally rounded to reduce current crowding effect [34-36]. Figure 4(c) shows the current-voltage characteristics of the device at an operation temperature of 2.4 K. The switching current $I_{SW}$ and the re-trapping current $I_r$ are 14.4 μA and 1.7 μA, respectively. Namely, the critical current density of our 60-nm-wide $Nb_{0.65}Ti_{0.35}N$ wires can be up to $3.45 \times 10^{10}$ A/m². Figure 4(d) presents the normalized intrinsic detection efficiency (IDE) at 1550 nm (solid squares) and dark count rate (DCR, hollow squares) of the detector as a function of bias current $I_b$. The detector shows saturated internal efficiency above 12 μA, with a saturation plateau approximately 2 μA. Due to the un-filtered direct fiber coupling configuration, the device shows a background DCR of 10 Hz, and is followed by an activated DCR at bias current above $I_b$~14 μA.

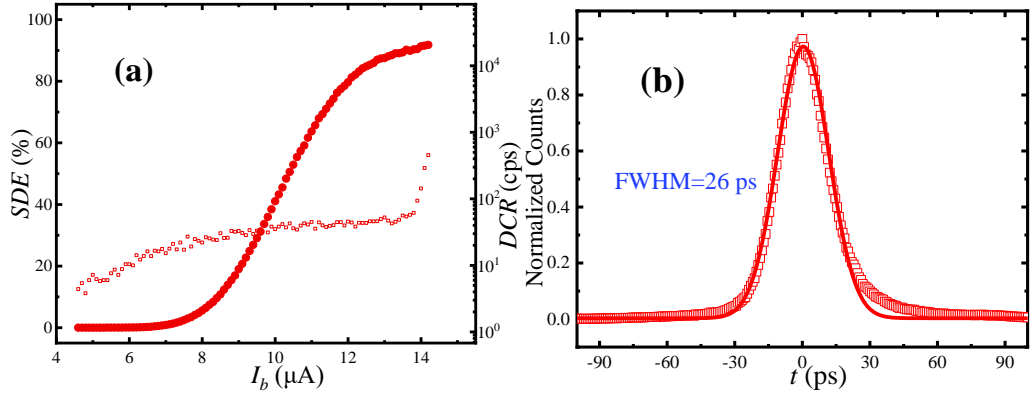

Figure 5 (a) System detection efficiency (solid squares) and dark count rate (open squares) as a function of bias current at 2.2 K. (b) The normalized photon counts dependence on the timing difference at a bias current of 13 μA for 1550 nm photons. The solid line indicates a fitting to a Gaussian distribution function.

Finally, we also deposited films and fabricated SNSPDs on the alternative $SiO_2/Ta_2O_5$ dielectric mirror, where the devices have the same sensitive area and nanowire width as that in Fig. 4. However, on the $SiO_2/Ta_2O_5$ dielectric mirror, the wire width of resulting superconducting nanowires is not as uniform as that on $SiO_2/Si$ substrates. The critical temperature of the resulting nanowire is consistent with those on thermally oxidized Si substrate. However, the switching current, $I_{SW} = 14.2$ μA, is slightly reduced as compared with the nanowires on $SiO_2/Si$ substrate. In figure 5(a), we show the bias current dependence of the system detection efficiency at 1550 nm and dark count rate, respectively, which is recorded at 2.4 K. Due to the variation of the wire width, the saturation plateau is then less constant as that on $SiO_2/Si$ substrates. The SDE reaches 90% at $I_b \sim 13.5$ μA, and it shows a maximal SDE of ≈ 92%. The intrinsic dark count becomes dominant above $I_b \sim 13.8$ μA. The timing jitter was also measured by using a time-correlated single-photon counting (TCSPC) method incorporated with a low temperature amplifier located at the 40 K stage [37]. Figure 5(b) shows the histogram of time-correlated photon counts at a bias current of 13 μA for 1550 nm photons. By fitting with a Gaussian distribution function, the timing jitter for the device is found to be around 26 ps.

**Conclusion**

In conclusion, a series of highly disordered $Nb_xTi_{1-x}N$ films were deposited by using magnetron sputtering, and the critical temperature of these films presented a dome like behavior on the Nb content. Based on the sheet resistance and residual resistivity ratio dependence on Nb content, we figured out the optimal $Nb_{0.65}Ti_{0.35}N$ films for SNSPDs fabrications. The fabricated $Nb_{0.65}Ti_{0.35}N$ SNSPDs showed a system detection efficiency as high as 92%. Moreover, the device presented a maximal bias current up to 14.2 μA, and a timing jitter as low as 26 ps at 2.4 K.

**Acknowledgements**

The work was supported financially by National Natural Foundation of China under Grant. No. 61971408 and 61827823, the Swedish Government Strategic Research Area in Materials

Science on Functional Materials at Linköping University (Faculty Grant SFO-Mat-LiU No. 2009 00971), the Knut and Alice Wallenberg foundation through the Wallenberg Academy Fellows program (KAW-2020.0196) and support for the Linköping Electron Microscopy Laboratory, and the Swedish Foundation for Strategic Research (SSF) support under the Research Infrastructure Fellow RIF 14‐0074. We acknowledge Robert Boyd for TEM operation. Daniel Primetzhofer from Uppsala University is acknowledged for Accelerator operation supported by Swedish Research Council VR-RFI (Contract No. 2019-00191) and the Swedish Foundation for Strategic Research (Contract No. RIF14-0053).